# Multi-year Long-term Load Forecast for Area Distribution Feeders based on Selective Sequence Learning

Ming Dong[a], Jian Shi[b], Qingxin Shi[c,*]

[a]: Dept. of Grid Reliability, Alberta Electric System Operator; T2P 0L4, AB, Canada
[b]: Dept. of Electrical and Computer Engineering, University of Houston; TX 77204-4005, USA
[c]: Dept. of Electrical Engineering and Computer Science, The University of Tennessee; TN 37996, USA
[*]: Corresponding author email: qshi1@vols.utk.edu

*Abstract -* Area feeder long-term load forecast (LTLF) is one of the most critical forecasting tasks in electric distribution utility companies. Cost effective system upgrades can only be planned out based on accurate feeder LTLF results. However, the commonly used top-down and bottom-up LTLF methods fail to combine area and feeder information and cannot effectively deal with component-level LTLF. The previous research effort on hybrid approach that aims to combine top-down and bottom-up approaches is very limited. The recent work only focuses on the forecast of the next one-year and uses a one-fit-all model for all area feeders. In response, this paper proposes a novel selective sequence learning method that can convert a multi-year LTLF problem to a multi-timestep sequence prediction problem. The model learns how to predict sequence values as well as the best-performing sequential configuration for each feeder. In addition, unsupervised learning is introduced to automatically group feeders based on load compositions ahead of learning to further enhance the performance. The proposed method was tested on an urban distribution system in Canada and compared with many conventional methods and the existing hybrid forecasting method. It achieves the best forecasting accuracy measured by three metrics AMAPE, RMSE and R-squared. It also proves the feasibility of applying sequence learning to multi-year component-level load forecast.

KEY WORDS: Long-term Load Forecast, Multi-timestep Sequence Prediction, Unsupervised learning

# 1. Introduction

For electric distribution utility companies, area feeder long-term load forecast (LTLF) is the task of forecasting all feeders' peak demand in a geographic area for the next few years. This problem is a component-level LTLF problem. The task is especially important because its results are used as the direct input for assessing the distribution system's power delivery capacity during normal operation as well as the restoration capability during contingencies. It is the foundation of distribution system planning work. Only based on accurate feeder load forecast results, distribution utility companies can plan long-term infrastructure upgrades or modifications in a cost-effective way [1-2].

In order to forecast the load growth in a planning area, most utility companies either employ top-down or bottom-up methods. However these methods fail to combine both area-level and component-level information together and hence often fail to accurately forecast component-level load such as all the distribution feeders that are located in the same area. Very limited effort has been put into researching hybrid methods that aim to combine the processes of top-down and bottom-up forecast. In response, this paper proposes a novel hybrid method that can effectively address this important forecasting problem.

This paper is organized as follows: in the beginning, it systematically reviews the previous top-down, bottom-up and hybrid methods used for LTLF. Based on detailed analysis and comparison, the paper proposes a new method that integrates the uses of unsupervised learning and selective sequence learning. A flowchart is given and following the flowchart, the paper discusses the unsupervised learning techniques for automatic feeder grouping. It then briefly introduces the area and feeder features to be used in the subsequent sequence learning step. In the sequence learning section, this paper systematically reviews the non-gated Recurrent Neural Network (RNN) and gated RNN based sequence learning models as well as three types of sequential configurations and their required dataset format. Section 7 explains the selective learning process for best-performing sequential configuration. In Section 8, the approach is applied to a large urban grid in Canada and discussed with case studies. Multiple conventional methods and the existing hybrid forecasting method are implemented on the same dataset and compared to the proposed method by using three performance metrics. The proposed method shows the best forecast accuracy and demonstrates the full feasibility of using sequence learning models for multi-year component-level load forecast.

# 2. Literature Review

In recent years, there have been a lot of research works on short-term load forecast (STLF) problems [3-9]. For example, [6] improves the forecasting accuracy of daily enterprise electricity consumption by using random forest and ensemble empirical mode decomposition; [7] proposes the method of using dynamic mode decomposition for short-term load forecast; [8] proposes the method of using a structure-calibrated support vector regression approach to forecast daily natural gas consumption; [9] proposes the method of using ensemble neuro-fuzzy model for short-term load forecast.

However, the research for component-level LTLF such as area feeder LTLF is rarely seen. According to [10], LTLF methods can be categorized into three approaches: top-down, bottom up and hybrid. The commonly used top-down and bottom-up LTLF approaches cannot effectively deal with component-level LTLF such as area feeder LTLF due to the following reasons:

- Top-down approach is based on area features such as area economy, demographics, weather and historical area loading. Previously, some methods directly apply univariate regression models to analyze historical area loading and its long-term trend [11-13]; some other methods apply multivariate regression models to analyze long-term relationship between area loading and other area features [14-17]. The results produced by these methods can reflect the area characteristics but cannot be directly applied to each component such as an individual feeder or transformer because strong variation exists at the component level. It is unrealistic to assume all components simply follow the area behaviour. As a result, for component-level LTLF, top-down approach is often subjectively used only to ensure that the component-level forecast results do not contradict with the area characteristics [10].

- Bottom-up approach is based on customer load information at the component level. The gathering of customer load information can be conducted through utility surveys, customer interviews and/or analyzing area development plans. For example, for a distribution feeder, major customers' long-term load information such as expected sizes of new loads, load maturation plan and/or long-term production plan can be collected, summarized and estimated to be yearly loading change and added to the current base feeder loading. Opposite to the top-down approach, this approach overly relies on the customer information and lacks the necessary understanding of long-term area behaviour. After all, it is very common for customers to overestimate or underestimate their load plans due to insufficient understanding of the area long-term economics.

As can be seen from above, top-down approach and bottom-up approach fail to combine area and feeder information and therefore cannot effectively use all the information for component-level LTLF. On the other hand, the research on hybrid forecasting approach that aims to combine top-down and bottom-up forecasting and overcome their drawbacks is very limited: [18] proposed a model to forecast individual household load that can be statistically adjusted by area features. For area feeder LTLF, [19] is the only found publication that made an attempt to combine the use of area features and feeder features. However, it has the following major drawbacks:

- It only focuses on the forecast of next one-year. However, in real application, planning engineers often need to forecast 3-5 years ahead. [19] did not propose a sound methodology and test the forecasting performance for multi-year forecast. It is therefore very limited in its real application value.

- It adopts a one-fit-all model for all area feeders regardless of the feeder type. However, in a planning area, there can be many different types of feeders such as residential-heavy feeders and commercial-heavy feeders. They can react to area economic and weather features in different ways. For example, a residential-heavy feeder is often less sensitive to economic conditions compared to feeders that have more commercial and industrial loads. Relying on one governing model for all feeders could lead to forecasting errors.

To address the above two major drawbacks, this paper made the following contributions, for the purpose of establishing a truly functional hybrid multi-year long-term load forecast method for area distribution feeders:

- It systematically studies different sequential configurations. Three types of sequential configurations were discussed: Single-year Recursive, Single-year with Interval and Multi-year Ahead. These sequential configurations can convert a multi-year LTLF problem to a multi-timestep sequence prediction problem and made multi-year component-level load forecast feasible.
- It proposes a novel selection mechanism that can learn, register and apply the best sequential configuration for each feeder in an area. This contribution can significantly improve the forecasting performance;
- It proposes the idea of applying unsupervised learning techniques to automatically group feeders into different groups first and then establish a set of sequence learning models for each group of feeders. This contribution reduces the accuracy loss caused by one-fit-all models.

The overall methodology of the proposed method is discussed in Section 3.

## 3. Overall Methodology

As discussed, the proposed method aims to effectively incorporate area and feeder level features in different historical years for the forecast of feeder peak demand in multiple years ahead. The structure of the proposed method is shown in Figure 1.

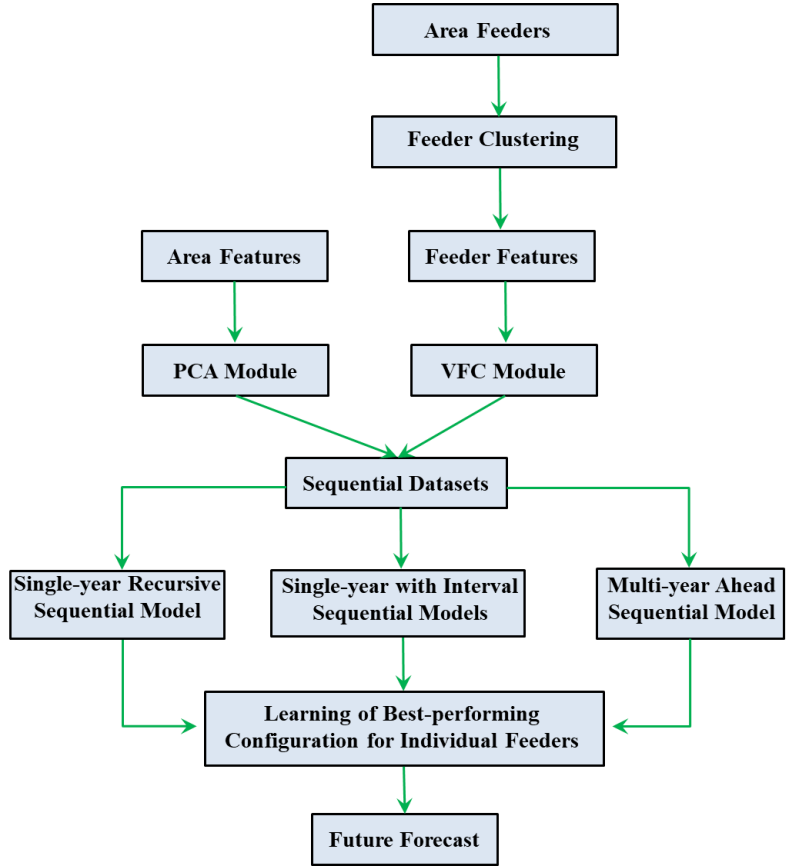

Figure 1. Workflow of the proposed method

In the beginning, all target feeders in a planning area are clustered into different groups by their load composition characteristics. Then feeder features are constructed and fed into the Virtual Feeder Conversion (VFC) module to eliminate data noises resulted from historical load transfer events [19]; in parallel, area economic and temperature features go through Principal Component Analysis (PCA) to reduce dimensions. In the end, processed area features and feeder features are combined to construct three different sequential datasets, corresponding to the Single-year Recursive, Single-year with Interval and Multi-year Ahead sequential configurations. For each cluster of feeders, models under the three sequential configurations are trained separately. Then for each feeder in the planning area, the best-performing sequential model is learned through a special evaluation mechanism discussed in Section 7 and gets registered for this feeder. Finally, for future forecast, three types of sequential configuration models trained for each cluster will be selected alternatively for different feeders to achieve the best overall forecasting accuracy.

## 4. Unsupervised Learning for Feeders

Different from supervised learning, unsupervised learning learns from data that is not pre-labeled [22]. It analyzes the commonalities between data points and groups similar data points together. In the proposed method, feeders in one area will be clustered based on the feeder load composition. Typically, each feeder contains residential, commercial and industrial loads. These three types of loads mix on a feeder according to certain percentages. Because each type of load responds to economy and temperature in different ways, our method first groups feeders with similar load compositions so that different prediction models can be established subsequently for each group of feeders. This step can enhance the prediction accuracy and is compared with treating all feeders in an area as only one group in Section 8.

K-Means clustering with Silhouette Analysis is chosen as the unsupervised learning method. K-Means clustering is a widely used method and has great efficiency and simplicity [23]. It requires only one input parameter *K* which is the expected number of clusters. To optimize the clustering performance, this paper further explains the use of Silhouette analysis as a clustering quality evaluation method to help select *K* [24].

*4.1 Feeder Load Composition*

Generally, there could be three types of loads on feeders: residential, commercial and industrial loads. Some feeders such as dedicated feeders may have only one type of loads while more commonly, many feeders contain more than one type. Feeder load composition can be described by the percentages of each type of loads. Residential load percentage *R*, commercial load percentage *C* and industrial load percentage *I* comply with:

$$R+C+I=1 \tag{1}$$

Therefore, to reduce the dimensionality and complexity of clustering, only two percentage numbers are required to characterize a feeder's composition. This means clustering can be performed on a 2-D basis. For example, we assume *R* and *C* are selected. *R* can be calculated by:

$$R = \frac{1}{N}\sum_{t=1}^{N}\sum_{i=1}^{n}\frac{L_{i,t}^{R}}{P_t} \times 100\% \tag{2}$$

where $P_t$ is the summer/winter peak load of the feeder in $t_{th}$ historical year; $L_{i,t}^R$ is the loading of residential load *i* at the feeder's peaking time in $t_{th}$ historical year; *n* is the total number of residential loads on this feeder; *N* is the number of historical years used for learning.

Similarly, commercial peak load percentage of a feeder is calculated by:

$$C = \frac{1}{N}\sum_{t=1}^{N}\sum_{i=1}^{n}\frac{L_{i,t}^{C}}{P_t} \times 100\% \tag{3}$$

where $L_{i,t}^C$ is the loading of commercial load $i$ at the feeder's peaking time in $t_{th}$ historical year; $n$ is the total number of commercial loads on this feeder.

After the above calculations, a feeder can be characterized with a vector $(R,C)$.

## 4.2 K-means Clustering

Mathematically, K-Means clustering is described as below: given a set of data points $(X_1, X_2, ..., X_n)$, where each data point is a q-dimensional real vector, K-Means clustering aims to group $l$ data points into $K\ (\leq l)$ clusters $S = \{S_1, S_2, ..., S_K\}$ so as to minimize the within-cluster variances of $S$. Formally, the objective function is defined as:

$$arg\ min \sum_{k=1}^{K} \sum_{x \in S_i} \|x - \mu_i\|^2 = arg\ min \sum_{k=1}^{K} |S_i|\ Var\ S_i \qquad (4)$$

where $\mu_i$ is the mean of data points in cluster $S_i$ [23]. The steps of K-Means are described in Algorithm 1.

To signify the numerical differences, the raw $R$ and $C$ percentage numbers can be further normalized using Min-Max normalization [22]:

$$R'_i = \frac{R_i - Min(R)}{Max(R) - Min(R)} \qquad (5)$$

where $R_i$ is the raw percentage number of residential load on $i_{th}$ feeder; $Min(R)$ and $Max(R)$ are the minimum and maximum values of all feeders' residential load percentage. Feature $C$ can be normalized in the same way.

---

**Algorithm 1: K-Means Clustering**

---

**Input**: $D = \{\mathbf{x_1}, \mathbf{x_2}, ... \mathbf{x_n}\}$  # dataset contains $n$ data points
  $K$ # expected number of clusters
**Output:** $K$ clusters
1: Randomly initialize $K$ centroids $\mathbf{C_1}$ to $\mathbf{C_k}$ for clusters $\mathbf{S_1}$ to $\mathbf{S_k}$
2: **while** stopping criterion not reached
3:   **for** $i \leftarrow 1$ **to** $n$
4:     Assign $\mathbf{x_i}$ to its nearest cluster $\mathbf{S}$ by measuring the distance between $\mathbf{x_i}$ and the centroid $\mathbf{C}$
5:   **end for**
6:   for $j \leftarrow 1$ **to** $K$
7:     $\mathbf{C_j} \leftarrow$ mean $(\mathbf{x} \in \mathbf{S_j})$
8:   **end for**
9: **end while**

The distance between any two feeders $F_1$ and $F_2$ can be calculated using standard Euclidean distance as below [22]:

$$d(F_1, F_2) = \sqrt{(R'_{F1} - R'_{F2})^2 + (C'_{F1} - C'_{F2})^2} \tag{6}$$

where $R'_{F1}$, $R'_{F2}$, $C'_{F1}$, $C'_{F2}$ are the normalized load composition features for two feeders $F_1$ and $F_2$.

*4.3 Clustering Quality Evaluation and Determination of Parameter K*

Silhouette analysis as a clustering quality evaluation method can be used to determine the optimal parameter $K$ from an initial range of $K$ values [24]. In this analysis, Silhouette coefficient $Q_r$ is used as an index to evaluate clustering quality. For a given data point $r \in S_r$, its $Q_r$ can be mathematically calculated following the steps below:

$$\begin{cases} Q_r = \dfrac{b_r - a_r}{max(a_r, b_r)} \\ a_r = \dfrac{1}{|S_r| - 1} \sum_{s \in S_r, r \neq s} d(r, s) \\ b_r = \min[\dfrac{1}{|S_v|} \sum_{v \in S_v} d(r, v)] \end{cases} \tag{7}$$

where $|S_r|$ is the number of members in cluster $S_r$ (i.e. cardinality); $S_v$ is any other cluster in the dataset; data point $v$ belongs to $S_v$; $d$ is the Euclidean distance between two data points measured by (6).

Equation (7) evaluates both the compactness and separation of produced clusters by K-means: for compactness, $a_r$ is the average distance of data point $r$ to all other points in the same cluster $S_r$. It reflects the intra-cluster compactness; $b_r$ is the smallest average distance of $r$ to all points in every other cluster that does not contain $r$. It reflects the inter-cluster separation. A big $b_r$ indicates a large inter-cluster separation seen from point $r$; in the end, $Q_r$ combines $a_r$ and $b_r$. A good intra-cluster compactness and inter-cluster separation together will results in a big $Q_r$.

(7) is the calculation for any single data point $r$. To evaluate the clustering quality of the entire dataset, average Silhouette coefficient is used and is given as below [24]:

$$Q_{avg} = \frac{1}{m} \sum_{i=1}^{m} Q_i \tag{8}$$

where $m$ is total number of data points in this dataset.

The steps of using Silhouette analysis to determine optimal cluster number $K$ are given as follows:

**Algorithm 2: Silhouette Analysis**

**Input**: $D=\{\mathbf{x_1}, \mathbf{x_2}, ... \mathbf{x_m}\}$  # dataset contains *m* data points
       $K \in \{\mathbf{1, 2, ... N}\}$  # Initial range for *K*
**Output:** Optimal cluster number $\boldsymbol{K_p}$
1: **for** $i \leftarrow 1$ **to** $N$
2: Apply K-means clustering (assuming *K=i)*
3: Calculate $\boldsymbol{Q_r}$ for each $\mathbf{x} \in D$
4: Calculate $\boldsymbol{Q_{avg}}$ for *D*
5: **end for**
6: $\boldsymbol{K_p} \leftarrow K$ with maximum $\boldsymbol{Q_{avg}}$

Through K-means clustering and Silhouette analysis, a number of feeders can be grouped automatically based on their load compositions. An example of clustering 300 feeders to 4 clusters is shown in Figure 2.

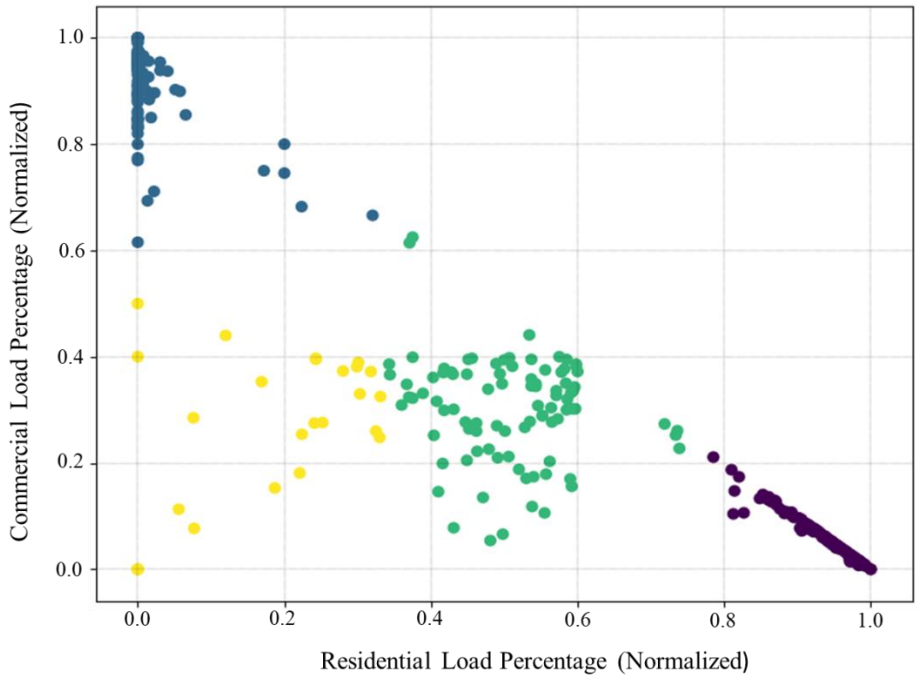

Figure 2. Example of clustering 300 feeders to 4 clusters by load composition features

## 5. Feature Selection and Processing

As a hybrid LTLF approach, both area features (top-down) and feeder features (bottom-up) are incorporated into modelling. By employing domain knowledge, useful raw features related to distribution feeder LTLF are selected. They need to be processed before fed into the sequence learning step.

*5.1 Area Features*

Area features describe the overall drivers in the planning area. Typical area features are listed in Table 1. The economic and demographic features have been explained in detail in the previous work [19]. "Extreme

Temperature Above Average" feature is the difference between maximum (summer)/minimum (winter) temperature of the current year and the historical average such as the 10-year average [25-26].

Table 1: Area Features

| Feature Name | Category |
|---|---|
| Real GDP Growth (%) | Economy |
| Total Employment Growth (%) | Economy |
| Industrial Production Index | Economy |
| Commodity Price | Economy |
| Population Growth (%) | Demographics |
| Net Migration | Demographics |
| Number of Housing Starts | Demographics |
| Extreme Temperature Above Average | Temperature |

*5.2 Feeder Load Features*

Feeder load features describe the detailed feeder-level load information.

- Major Customer Net Load Change: this feature is the estimated net load change of all major customers on the feeder between any two years. Utility companies usually have special teams that conduct surveys, interview customers, study developer/city development plans or rely on external consultants to collect such information. The aggregated net change is the summation of all estimated load changes from major customers on the feeder. It should be noted that this paper focuses on the forecasting method itself and treats such information as given input for the discussed methodology.

- Distributed Energy Resource (DER) and Electrical Vehicle (EV) Adoption Change: similar to Major Customer Net Load Change, DER and EV can be considered in certain planning areas with high concentrations. Future DER and EV annual adoption can be forecasted in separate tasks [27-28]. In this paper, they are treated as given input for the discussed methodology.

- Base Peak Demand: the current summer or winter peak demand is used as the base demand. It provides a baseline while most of other area and feeder load features focus on the change from the current year to the forecast year.

*5.3 Principal Component Analysis for Area Features*

Many economic and demograhic area features are highly correlated. To improve the prediction accuracy, PCA shall be applied to reduce the dimensionality. This process has been explained in detail in [19].

*5.4 Virtual Feeder Conversion for Feeder Features*

When dealing with long-term historical feeder loading data, it is inevitable to encounter load transfer events which can suddenly disrupt the original trend of feeder loading and introduce interference. [19]

proposed the idea of combining two or more feeders with load transfer events to one virtual feeder so that the transfers between them can be ignored. This processing technique can effectively eliminate the data noise caused by load transfers and continues to be used in this research.

## 6. Sequence Learning for LTLF

This section briefly reviews the theory of sequence learning models. We start from non-gated RNN and go on to explain gated RNN. Three different sequential configurations as well as their dataset formats are also discussed.

*6.1 Non-gated RNN*

As shown in Figure 3, a RNN is a group of feed-forward neural networks (FNN) connected in series. Hidden neurons of the FNN at a previous time step are connected with the hidden neurons of the FNN at the following time step. This can make hidden state at the last time step $H_{t-1}$ pass into the current time step. $H_{t-1}$ is then combined with the current input $X_t$ to produce the current hidden state $H_t$ through trained weights $W_h$ and $W_x$. This process continues to the next time step until the end of the sequence. In this unique way, RNN is able to make use of historical information and does not treat one time step as an isolated point. This made RNN suitable for forecasting tasks such as word prediction and load forecast where the output of current time step is not only related to the current input but also previous time steps. An unfolded RNN structure is shown in Figure 3.

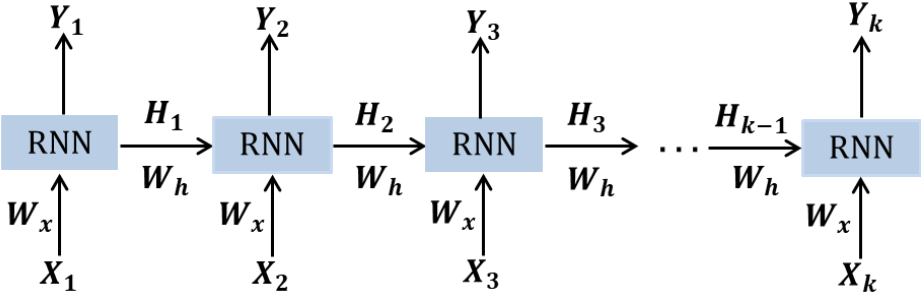

Figure 3. Illustration of an unfolded RNN structure

In spite of the obvious advantages, the training of non-gated RNN can be unstable due to an intrinsic problem called vanishing/exploding gradient [29-30]. During back propagation of RNN, gradient value may become too small to drive the network update or too large to stabilize the training. This problem leads to the invention of gated RNNs which successfully solve the vanishing/exploding gradient problem through sophisticated gate controls [30-31]. In recent years, gated RNNs have replaced non-gated RNN as the industry standard.

## 6.2 Gated RNN

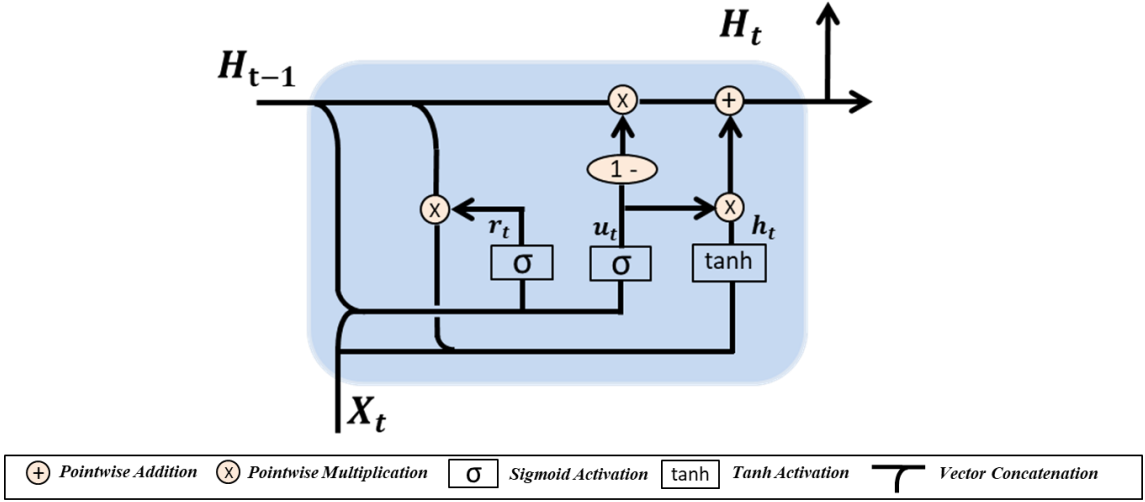

Figure 4. A GRU unit diagram

Gated Recurrent Unit (GRU) network and Long Short-term Memory (LSTM) are two commonly used gated RNNs. As suggested in [33-36], GRU has a performance similar to LSTM but is much faster. Hence, in this paper, we only adopt GRU as the sequence learning model. In addition to the chain-structure of non-gated RNN, a GRU network has a GRU unit diagram shown in Figure 4.

An element of the reset gate $r_t$ is calculated by:

$$r_t = \sigma(W_r \cdot [H_{t-1}, X_t] + b_r) \tag{9}$$

where $[H_{t-1}, X_t]$ is the concatenated vector of hidden state vector $H_{t-1}$ at the previous time step and the input vector $X_t$ at the current time step; $W_r$ and $b_r$ are the weight vector and bias.

An element of the update gate $u_t$ is calculated in a very similar way, with different weight vector $W_u$ and bias $b_u$:

$$u_t = \sigma(W_u \cdot [H_{t-1}, X_t] + b_u) \tag{10}$$

According to the information flow illustrated in Figure 4, a temporary value $h_t$ is calculated by:

$$h_t = tanh(W_h \cdot [r_t \otimes H_{t-1}, X_t] + b_h) \tag{11}$$

In the end of the information flow, the hidden state vector $H_t$ at the current time step $t$ is generated by using hidden state vector $H_{t-1}$, update gate vector $u_t$ and temporary vector $h_t$ through pointwise multiplication and addition:

$$H_t = (1 - u_t) \otimes H_{t-1} + u_t \otimes h_t \tag{12}$$

Through (9)-(12), hidden state $H_t$ is updated from one time step to the next until the end of the sequence.

*6.3 Sequential Configurations*

A RNN can be implemented in different sequential configurations [30-31]. For the multi-year forecast problem discussed in this paper, there are three suitable sequential configurations: Single-year Recursive, Single-year with Interval and Multi-year Ahead. A 3-year load forecast is used as an example in here for illustration purpose. The schematics of the three sequential configurations are shown in Figure 5.

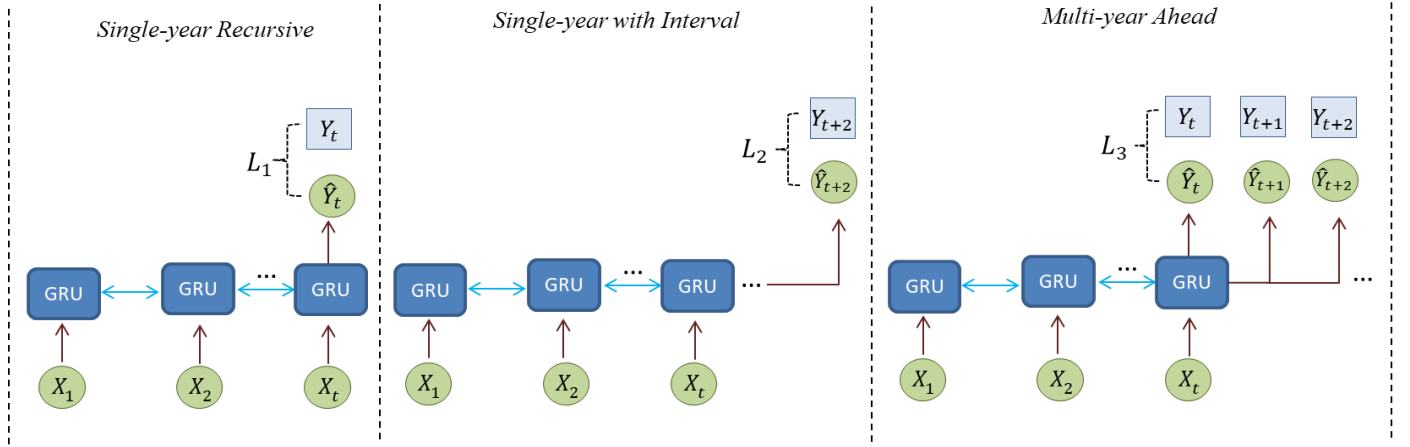

Figure 5. Three different sequential configurations

The theoretical differences of these configurations lie in the network loss functions. For all neural networks, network loss is converted to gradient and drives the training of neural network through back propagation and mathematic chain-rule [22]. Here, Mean Absolute Error (MAE) is chosen to calculate the network loss.

The Single-year Recursive configuration has been used in [19] and only forecasts the next one-year. This means in order to forecast multiple years ahead, for example the next 3 years, the model has to be used recursively: the first forecast year $Y_t$ is firstly forecasted and then $Y_t$ is used as a known input to forecast the second forecast year $Y_{t+1}$; similarly, $Y_{t+1}$ is then used to forecast the third forecast year $Y_{t+2}$. Although this configuration has a good performance for forecasting the first forecast year, the error may become larger for later years especially when applying this configuration for a longer forecast window. This is because the forecasted values are recursively used as input and the error can get accumulated as the forecast moves into later years. This issue becomes more prominent if the feature fluctuation between neighbouring years is large. The loss function $L_1$ of Single-year Recursive configuration is:

$$L_1 = \frac{1}{n} \sum_{j=1}^{n} |Y_t^j - \hat{Y}_t^j| \tag{13}$$

where $n$ is the training batch size; $Y_t^j$ is the first forecast year $t$'s actual peak demand in $j_{th}$ record in the training batch; $\hat{Y}_t^j$ is the first forecast year $t$'s forecasted peak demand in $j_{th}$ record in the training batch.

Different from the Single-year Recursive configurations, the Single-year with Interval configuration directly forecasts a specific future year in the forecast window. When it is used to forecast the first forecast year, it is equivalent to applying the Single-year Recursive configuration because the yearly interval is zero. However, it becomes different for forecasting later years in the forecast window: for later years, the area and feeder features between the forecast year and the current year are summed up to reflect the total change over the interval years. Then the forecast year is directly forecasted. Compared to Single-year Recursive configuration, applying this configuration is not a recursive process and can reduce error in many cases. The trade-off is that for a forecast window of $T$ years, $T$ models need to be trained to forecast every different year in the forecast window. The loss function $L_2$ for Single-year with Interval configuration is:

$$L_2 = \frac{1}{n} \sum_{j=1}^{n} |Y_{t+f-1}^j - \hat{Y}_{t+f-1}^j| \tag{14}$$

where $n$ is the training batch size; $Y_{t+f-1}^j$ is the $f_{th}$ forecast year's actual peak demand in $j_{th}$ record in the training batch; $\hat{Y}_{t+f-1}^j$ is the $f_{th}$ forecast year's forecasted peak demand in $j_{th}$ record in the training batch. The example in Figure 5 shows the scenario when $f = 2$ (i.e. forecasting the third forecast year); when $f = 1$ (i.e. forecasting the first forecast year), (14) is equivalent to (13).

Compared to the first two configurations, the Multi-year Ahead configuration outputs the results of multiple years in the forecast window all at once. The advantage is its great efficiency and emphasis on the holistic accuracy of the forecast window. On the other hand, it does not utilize the forecasted area and feeder features in all future years (after the first forecast year). Using less information is not necessarily bad when the accuracy of some information such as customer load information or far-out economic forecast cannot be guaranteed. Its loss function $L_3$ is:

$$L_3 = \frac{1}{nT} \sum_{j=1}^{n} \sum_{i=t}^{t+T-1} |Y_i^j - \hat{Y}_i^j| \tag{15}$$

where $n$ is the training batch size; $Y_i^j$ is $i_{th}$ year's actual peak demand in $j_{th}$ record in the training batch; $\hat{Y}_i^j$ is $i_{th}$ year's forecasted peak demand in $j_{th}$ record in the training batch; $T$ is the length of forecast window in number of years. Different from $L_1$ and $L_2$, $L_3$ measures the average error of the entire forecast window.

*6.4 Sequential Datasets*

To fit into the above three RNN sequential configurations, data records must be grouped by a fixed number of time steps following specific formats. This is different from traditional single-row datasets that are commonly used for other types of supervised learning methods. Table 2 to Table 4 are examples for Single-year Recursive, Single-year with Interval and Multi-year Ahead configurations.

Table 2 shows a dataset example for Single-year Recursive configuration. Data record ID 21 is taken as an example for explanation: in this record, the goal focuses on the forecast of 2011's peak load by using the previous-years' base peak demand in 2008, 2009 and 2010 as well as the yearly economic and temperature features in 2009, 2010 and 2011. EP1 and EP2 are the processed economic-population growth features after applying PCA. ETAA is the "Extreme Temperature Above Average" feature and MCNLC is the "Major Customer Net Load Change" feature as mentioned in Section 5. The third year 2011 is the forecast year and its actual peak demand is also included in the training record.

Table 2: Dataset Example for Single-year Recursive Configuration

| Data Record ID | Feeder ID | Input Year | Base Peak Demand | Yearly Features | | | | Actual Peak Demand (Forecast Year) |
| | | | | EP1 | EP2 | ETAA | MCNLC | |
|---|---|---|---|---|---|---|---|---|
| … | … | … | … | … | … | … | … | … |
| 21 | 0050 | 2009 | 433 A | -0.64 | 0.44 | 0.7°C | 42 A | 550 A (2011) |
| | 0050 | 2010 | 502 A | -0.16 | 0.31 | -1.3°C | 34 A | |
| | 0050 | 2011 | 554 A | 0.33 | -0.31 | 3.4°C | 0 A | |
| 22 | 0050 | 2010 | 502 A | -0.16 | 0.31 | -1.3°C | 34 A | 521 A (2012) |
| | 0050 | 2011 | 554 A | 0.33 | -0.31 | 3.4°C | 0 A | |
| | 0050 | 2012 | 550 A | -0.06 | -0.17 | -2.2°C | -21 A | |
| … | … | … | … | … | … | … | … | … |

Similarly, Table 3 shows a dataset example for Single-year with Interval configuration. Different from Single-year Recursive configuration, the two data records aim to forecast 2012 and 2013 instead of the last input year 2011; Table 4 shows a dataset example for Multi-year Ahead configuration. Taking data record ID 35 as an example, the goal is not only to forecast 2011's peak load but also the peak load in 2012 and 2013. The loss function takes all three years' errors into consideration and is therefore less biased if 2011's loading experienced an unusual change. Compared to Single-year Recursive configuration, its focus on year 2011 is weaker as the goal is not only to forecast 2011 but also 2012 and 2013 all at once.

Table 3: Dataset Example for Single-year with Interval Configuration

| Data Record ID | Feeder ID | Input Year | Base Peak Demand | Yearly Features | | | | Actual Peak Demand (Forecast Year) |
|---|---|---|---|---|---|---|---|---|
| | | | | EP1 | EP2 | ETAA | MCNLC | |
| … | … | … | … | … | … | … | … | … |
| 67 | 0050 | 2009 | 433 A | -0.64 | 0.44 | 0.7°C | 42 A | 521 A (2012) |
| | 0050 | 2010 | 502 A | -0.16 | 0.31 | -1.3°C | 34 A | |
| | 0050 | 2011 | 554 A | 0.29 | -0.50 | -2.2°C | -21 A | |
| 68 | 0050 | 2009 | 433 A | -0.64 | 0.44 | 0.7°C | 42 A | 537 A (2013) |
| | 0050 | 2010 | 502 A | -0.16 | 0.31 | -1.3°C | 34 A | |
| | 0050 | 2011 | 554 A | 0.07 | 0.28 | 1.8°C | 20 A | |
| … | … | … | … | … | … | … | … | … |

Table 4: Dataset Example for Multi-year Ahead Configuration

| Data Record ID | Feeder ID | Input Year | Base Peak Demand | Yearly Features | | | | Actual Peak Demand (Forecast Year) |
|---|---|---|---|---|---|---|---|---|
| | | | | EP1 | EP2 | ETAA | MCNLC | |
| … | … | … | … | … | … | … | … | … |
| 35 | 0050 | 2009 | 433 A | -0.64 | 0.44 | 0.7°C | 42 A | 550 A (2011) |
| | 0050 | 2010 | 502 A | -0.16 | 0.31 | -1.3°C | 34 A | 521 A (2012) |
| | 0050 | 2011 | 554 A | 0.33 | -0.31 | 3.4°C | 0 A | 537 A (2013) |
| 36 | 0050 | 2010 | 502 A | -0.16 | 0.31 | -1.3°C | 34 A | 521 A (2012) |
| | 0050 | 2011 | 554 A | 0.33 | -0.31 | 3.4°C | 0 A | 537 A (2013) |
| | 0050 | 2012 | 550 A | -0.06 | -0.17 | -2.2°C | -21 A | 549 A (2014) |
| … | … | … | … | … | … | … | … | … |

## 7. Learning of Best-performing Sequential Configuration

For different feeders, the three sequential configurations presented in Section 6 may have different performances depending on the input information. For example, the Single-year Recursive configuration places all emphasis on the next one-year and can be accurate if the external drivers and customer load information (including assumed values for future years) are accurate; Single-year with Interval configuration can be effective when dealing with feeders with large yearly fluctuations as it reduces the error from the recursive process; Multi-year Ahead configuration can be more accurate for saturated feeders which are less sensitive to external drivers or feeders with less accurate customer load information.

However, it is challenging to manually analyze each feeder's characteristics and picks the best-performing sequential configuration for each feeder. This paper proposes an automatic way of learning the best-performing Sequential Configuration for individual feeders based on historical data. Figure 6 shows a training set example with 20-year data. The goal is to produce a three-year window forecast. This window slides from the fourth year to eighteenth year and for each slide, a MAE is calculated for each sequential configuration. Therefore, it produces in total 15 MAEs for performance evaluation.

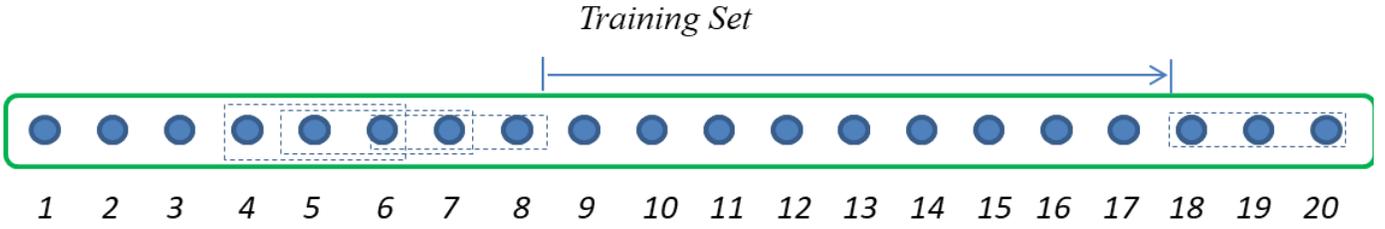

Figure 6. Sliding window over training set for Sequential Configuration learning

Mathematically, the performance index for each sequential configuration is:

$$P = \frac{1}{N - 2T' + 1} \sum_{j=1}^{N-2T'+1} \sum_{i=1}^{T'} |Y_i^j - \hat{Y}_i^j| \tag{16}$$

where $N$ is the total number of years in training set; $T'$ is the forecast window length; $Y_i^j$ is the actual peak demand of $i_{th}$ year in the $j_{th}$ window; $\hat{Y}_i^j$ is the forecasted peak demand of $i_{th}$ year in the $j_{th}$ window.

For each feeder, (16) is used to calculate performance index $P$ for all three sequential configurations. The sequential configuration with the smallest $P$ index is then selected and registered for this particular feeder. This configuration will be used on test set and future forecast for the feeder.

## 8. Validation and Application

The discussed method was applied to a large urban distribution system (City of Calgary) in Canada to forecast a three-year window peak demand in both summer and winter. In total 403 distribution feeders and their past 18-year annual data (2001-2018) were used to create the dataset. 2001-2015 data were used to form the training set and 2016-2018 three-year data were used to form the test set. Feeders were clustered into 4 groups using the proposed clustering techniques. For each sequential configuration, a 5-neuron input layer for 5 features exemplified in Table 2-4 is used; a 10-neuron GRU hidden layer is determined through grid search method. For the output layer, Single-year Recursive and Single-year with Interval configurations use 1 neuron while Multi-year Ahead configuration uses 3 neurons. All neurons use ReLU activation function. A 20% neuron dropout rate is also applied to avoid model overfitting. Since Mean Absolute Percent Error (MAPE) is a widely used metric for forecasting tasks [37-38], we use the average of MAPE (AMAPE) in the three test years 2016-2018 to evaluate the forecast performance for each feeder. Mathematically, it is given by:

$$AMAPE = \frac{1}{3m} \sum_{j=1}^{3} \sum_{i=1}^{m} \left| \frac{Y_i^j - \hat{Y}_i^j}{\hat{Y}_i^j} \right| \times 100\% \tag{17}$$

where $m$ is the number of feeders in the planning area; $\hat{Y}_i^j$ is the actual peak demand of $i_{th}$ feeder in $j_{th}$ year (starting in 2016); $Y_i^j$ is the forecasted peak demand of $i_{th}$ feeder in $j_{th}$ year.

In addition to AMAPE, two auxiliary metrics RMSE and $R^2$ (R-squared) are also used for performance measurement [22,38]. Mathematically, they are given as below

$$RMSE = \sqrt{\sum_{i=1}^{3m} \frac{(Y_i - \hat{Y}_i)^2}{3m}} \tag{18}$$

$$R^2 = 1 - \frac{\sum_{i=1}^{3m}(\hat{Y}_i - Y_i)^2}{\sum_{i=1}^{3m}(\hat{Y}_i - \bar{Y}_i)^2} \tag{19}$$

where $\hat{Y}_i$ is the actual peak demand of $i_{th}$ testing record; $Y_i$ is the predicted value corresponding to $\hat{Y}_i$; $\bar{Y}_i$ is the average actual peak demand of all testing records; $3m$ is the total number of testing records in three years.

When $R^2 = 1$, the prediction matches exactly with the actual every time in the test; when $R^2 = 0$, the performance is equivalent to always using the average of all actual values as prediction every time in the test; if the prediction accuracy is even worse than blindly using the average of actual values, $R^2$ can become a negative value. Different from AMAPE, RMSE and $R^2$ use absolute errors instead of percentage error for performance measurement. As a result, their values can be affected by the absolute magnitude of power demand and may not always align with AMAPE in comparison. This phenomenon has been observed in Section 8.4.

*8.1 Effect of Selective Sequence Learning*

To test the effect of the proposed selective sequence learning (SSL) method, the proposed method is compared to using only single sequential configurations for all feeders in summer and winter. The results are shown in Figure 7.

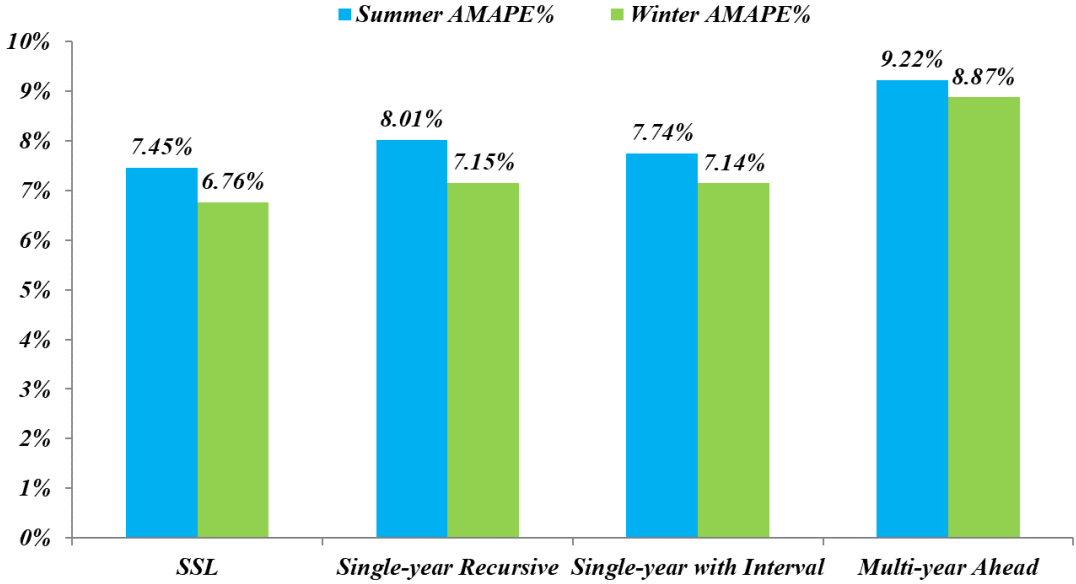

Figure 7. AMAPE comparison with using single sequential configuration

As can be seen, the selective sequence learning method demonstrates a lower AMAPE against all single sequential configurations. It is expected because by using SSL, each feeder now learns the best-performing sequential configuration during the training process and uses it for forecast. Performances measured by RMSE and $R^2$ also follow the same trend and are summarized in Table 5.

Table 5: Performance Comparison with using Single Sequential Configurations

| Season | Performance Metric | SSL | Single-year Recursive | Single-year with Interval | Multi-year Ahead |
|---|---|---|---|---|---|
| Summer | AMAPE | 7.45% | 8.01% | 7.74% | 9.22% |
| | RMSE | 30.5 | 36.1 | 33.2 | 42.7 |
| | $R^2$ | 0.916 | 0.894 | 0.903 | 0.859 |
| Winter | AMAPE | 6.76% | 7.15% | 7.14% | 8.87% |
| | RMSE | 29.5 | 33.2 | 32.8 | 38.4 |
| | $R^2$ | 0.934 | 0.908 | 0.922 | 0.868 |

*8.2 Effect of Feeder Clustering*

Another contribution of this paper is the proposed feeder clustering step. This step groups feeders with similar load compositions together. The rationale behind this is that feeders with similar load compositions respond to external drivers in similar ways and therefore have similar growth patterns. Therefore, if separate learning can be established for separate feeder groups, better performance could be achieved. This is proven by comparing AMAPE after using feeder clustering and AMAPE without using feeder clustering (treating all area feeders as one group and apply the same set of sequential models). The results are shown in Figure 8.

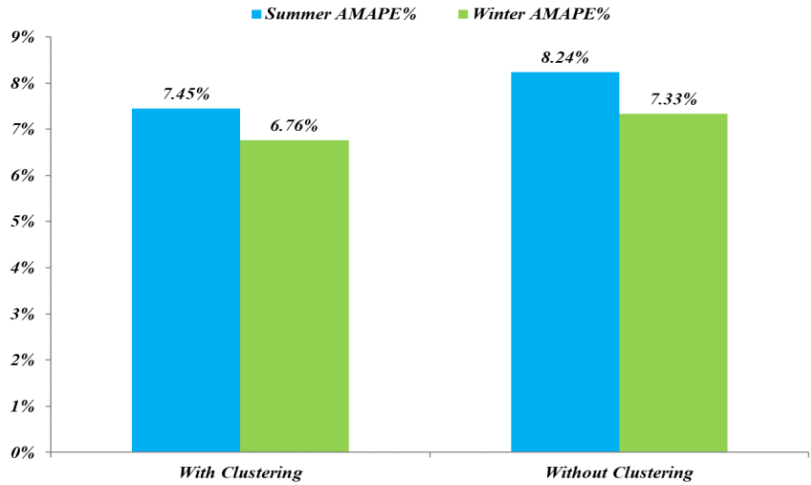

Figure 8. AMAPE comparison with and without feeder clustering

Performances measured by RMSE and $R^2$ also follow the same trend and are summarized in Table 6.

Table 6: Performance Comparison with and without Feeder Clustering

| Season | Performance Metric | With Clutering | Without Clustering |
|---|---|---|---|
| Summer | AMAPE | 7.45% | 8.24% |
|  | RMSE | 30.5 | 37.9 |
|  | $R^2$ | 0.916 | 0.887 |
| Winter | AMAPE | 6.76% | 7.33% |
|  | RMSE | 29.5 | 33.8 |
|  | $R^2$ | 0.934 | 0.894 |

*8.3 Comparison with Conventional Models*

In the end, the proposed model was compared to various other models specified as below. To ensure fairness, the same feature processing techniques discussed in Section 5 are also used for conventional models. The hyper parameters for hidden layers used in the neural network based models are tuned through grid search method.

- Bottom-up model: as discussed in Section 2, this is the method to build a feeder load forecast based on major customer load information. Mathematically, the "Major Customer Net Load Change" feature from 2016 to 2018 was gathered and added to the previous year's peak demand recursively to approximately estimate the following year's peak demand.

- ARIMA model: for each feeder, its peak demand data between 2001 and 2015 was fed into an ARIMA model for training. ARIMA (2,0,0) was chosen after experiments because it gave the best forecast result among different ARIMA order parameters for the dataset. Then the peak demand values between 2016 and 2018 were calculated recursively.

- One-input year recursive FNN (ORF): for each feeder, only one-input year's features are used for forecast. The features are the same as used in the proposed method, i.e. "Previous-Year Peak Demand", EP1, EP2, "Extreme Temperature over Average" and "Major Customer Net Load Change". A traditional FNN model is used, with a 5-neuron input layer, two 6-neuron hidden layers and a 1-neuron output layer. Peak demand values between 2016 and 2018 were forecasted recursively.

- Three-input year recursive FNN (TRF): a traditional FNN model is used to incorporate all the features from three input years (in total 15 features) to forecast the third year's peak demand. The FNN has a 15-neuron input layer, two 10-neuron hidden layers and a 1-neuron output layer. Peak demand values between 2016 and 2018 were forecasted recursively. The main difference between this method and the proposed many-to-one sequence prediction is that it uses the FNN structure instead of RNN structure.

- Three-input year non-recursive FNN (TNF): a traditional FNN model is used to incorporate all the features from three input years (in total 15 features) to forecast the three forecast years' peak demand. The FNN has a 15-neuron input layer, two 12-neuron hidden layers and a 3-neuron output layer representing peak demands in three years. Peak demand values between 2016 and 2018 were forecasted all at the same time. The main difference between this method and the proposed shifted many-to-many sequence prediction is that it uses the FNN structure instead of RNN structure.

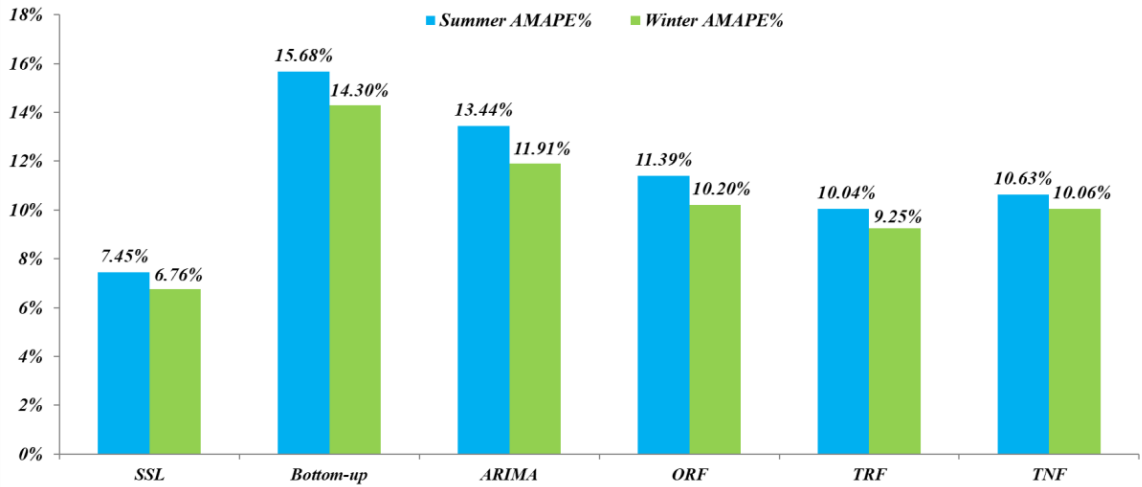

Figure 9. AMAPE comparison with conventional models

The comparison is summarized in Figure 9. As can be seen, the proposed method outperforms all 5 non-sequence prediction based conventional methods. Performances measured by RMSE and $R^2$ also follow the same trend and are summarized in Table 7.

Table 7: Performance Comparison with Conventional Models

| Season | Performance Metric | SSL | Bottom-up | ARIMA | ORF | TRF | TNF |
|---|---|---|---|---|---|---|---|
| Summer | AMAPE | 7.45% | 15.68% | 13.44% | 11.39% | 10.04% | 10.63% |
| | RMSE | 30.5 | 66.5 | 58.8 | 54.1 | 44.6 | 46.3 |
| | $R^2$ | 0.916 | 0.616 | 0.745 | 0.772 | 0.830 | 0.796 |
| Winter | AMAPE | 6.76% | 14.30% | 11.91% | 10.20% | 9.25% | 10.06% |
| | RMSE | 29.5 | 63.4 | 50.6 | 48.7 | 42.1 | 45.2 |
| | $R^2$ | 0.934 | 0.687 | 0.756 | 0.793 | 0.859 | 0.848 |

The superior performance of the proposed method (SSL) against other methods can be understood from four perspectives: it effectively uses both area features and feeder features. In comparison, the bottom-up and ARIMA methods can only use one aspect of information; it effectively use multiple historical years of data for forecast. In comparison, ORF can only use one-year data; it uses a GRU based sequential learning model. In comparison, TRF and TNF are based on regular FNN models and this network structure cannot analyze the relationship of different years (timesteps); furthermore, the proposed method uses unsupervised learning to pre-classify feeder types and selects the best sequential configuration for each feeder.

*8.4 Performance in Each Cluster*

As discussed above, the 403 feeders in the planning area were grouped into 4 clusters by load composition. This section discusses the testing results for each cluster. The number of members in each cluster, the load composition features of each cluster's centroid and the 2016-2018 average loadings are provided in Table 8.

Table 8: Cluster Composition

| Feeder Cluster ID | $(R, C)$ of Cluster Centroid | Number of Cluster Members | Average Summer Loading (A) | Average Winter Loading (A) |
|---|---|---|---|---|
| 1 | (77%,15%) | 122 | 338 | 386 |
| 2 | (30%,48%) | 97 | 271 | 303 |
| 3 | (21%,19%) | 48 | 374 | 392 |
| 4 | (27%,31%) | 136 | 270 | 319 |

As can be seen from the second column in Table 8, the feeders in Cluster 1 have significantly more residential load than commercial and industrial load; the feeders in Cluster 2 have more commercial load than residential and industrial load; the feeders in Cluster 3 have significantly more industrial load than residential and commercial load; the feeders in Cluster 4 have more balanced load types than in the other three clusters, with comparable amount of residential, commercial and industrial load. The summer and winter forecasting accuracies of these clusters measured by AMAPE are presented in Figure 10. It is shown that Cluster 3 has relatively higher forecasting error. This is probably because Cluster 3 is industrial load heavy and industrial load can fluctuate more drastically between years and operate in a more random pattern.

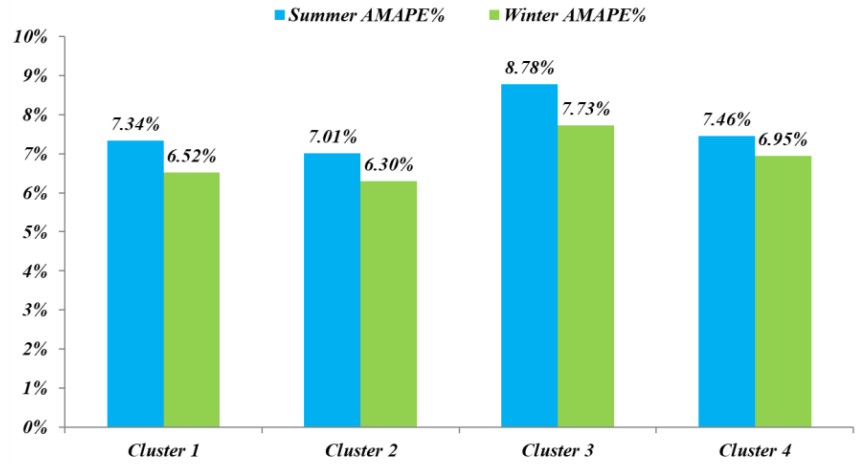

Figure 10. AMAPE comparison of different feeder clusters

Table 9: Performance Comparison with Different Clusters

| Season | Performance Metric | Cluster 1 | Cluster 2 | Cluster 3 | Cluster 4 |
|---|---|---|---|---|---|
| Summer | AMAPE | 7.34% | 7.01% | 8.78% | 7.46% |
| | RMSE | 31.3 | 23.1 | 41.2 | 28.2 |
| | $R^2$ | 0.829 | 0.870 | 0.708 | 0.840 |
| Winter | AMAPE | 6.52% | 6.30% | 7.73% | 6.95% |
| | RMSE | 30.1 | 21.7 | 38.1 | 28.6 |
| | $R^2$ | 0.850 | 0.929 | 0.783 | 0.862 |

However, it is noticed that RMSE and $R^2$ in Table 9 do not completely align with the observed trend of AMAPE between different clusters. The exception is on Cluster 1 and Cluster 4: although Cluster 4's AMAPE is slightly higher than Cluster 1, its performance measured by RMSE and $R^2$ is slightly better than Cluster 1. This is because Cluster 4's average loading is lower than Cluster 1 and results in smaller absolute errors. Since RMSE and $R^2$ use absolute errors as shown in equation (18) and (19), when the AMAPE difference between two clusters is small but the average loading difference is large, the absolute error based comparison may not necessarily align with percentage based comparison (AMAPE).

Furthermore, the numbers of sequential configurations registered within each cluster are checked and summarized in Table 10.

Table 10: Best-performing Sequential Configurations Selected in Each Cluster

| Feeder Cluster ID | Number of Cluster Members | Single-year Recursive | Single-year with Interval | Multi-year Ahead |
|---|---|---|---|---|
| 1 | 122 | 5 | 66 | 51 |
| 2 | 97 | 14 | 44 | 39 |
| 3 | 48 | 6 | 40 | 2 |
| 4 | 136 | 48 | 68 | 20 |
| Total | 403 | 73 | 218 | 112 |

As can be seen from Table 10, overall the least selected best-performing sequential configuration is the Single-year Recursive Configuration. This is within expectation because this configuration suffers from the recursive error for multi-year forecast; the most selected best-performing sequential configuration is the Single-year with Interval Configuration. It is significantly better than the other two configurations, especially for Cluster 3 when the load fluctuation between years is large due to its heavy industry load presence; the Multi-year Ahead Configuration performs better when residential load is heavier in the composition. This could be due to the fact that residential load is often more stable and less sensitive to external economic drivers in later years.

*8.5 Error Progression in Forecast Window*

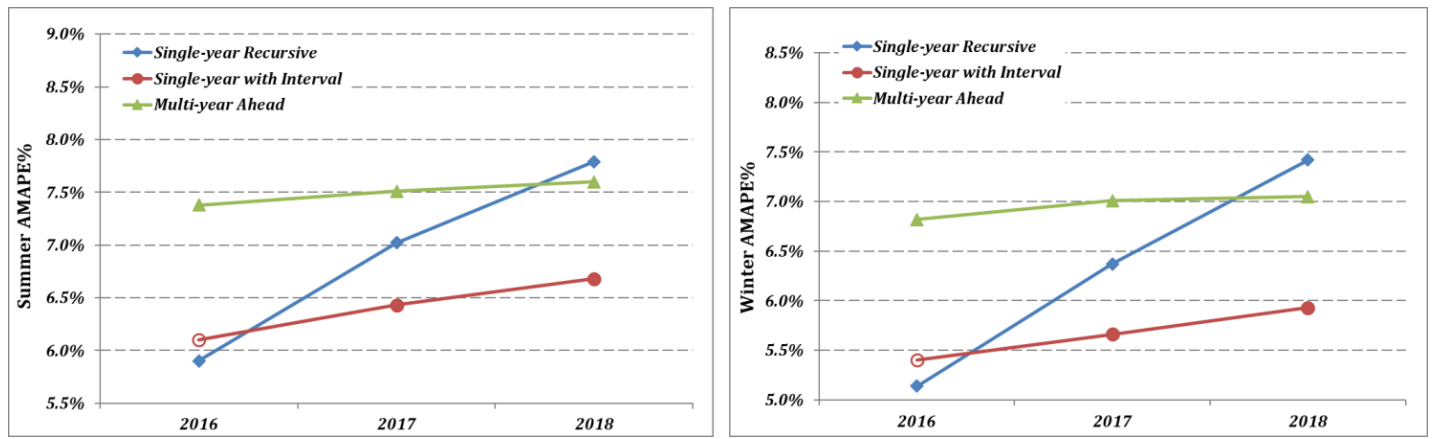

Figure 11. Performance comparison of different feeder clusters

The proposed method has been applied to the 403 feeders in the area in the testing period 2016-2018. Figure 11 shows the AMAPE progression over the three-year forecast window for the feeder groups under each best-performing configuration indicated in Table 6. It is found that the AMAPE for Single-year Recursive configuration increases quickly in year 2 and year 3. This is again due to the recursive nature of this configuration. In comparison, the other two configurations are much more stable. The AMAPE of Single-year with Interval configuration only rises slightly because the area economic forecast itself is less accurate in the later years of the forecast window. The AMAPE of Multi-year Ahead configuration is quite stable because it is trained based on the overall accuracy of all three years in the forecast window and does not rely on area features in the later years of the forecast window.

*8.6 Application Resutls For Next 5 Years*

In the end, the above models established based on historical data are applied to forecast the feeder peaks of the next 5 years (2019 to 2023) in the same area. The average feeder peaks in 4 clusters for the next 5 years in summer and winter are presented in Figure 12. All 4 clusters of feeders experience load growth in the next 5 years. Table 11 summarizes the average growth rates of the 5 years in both summer and winter.

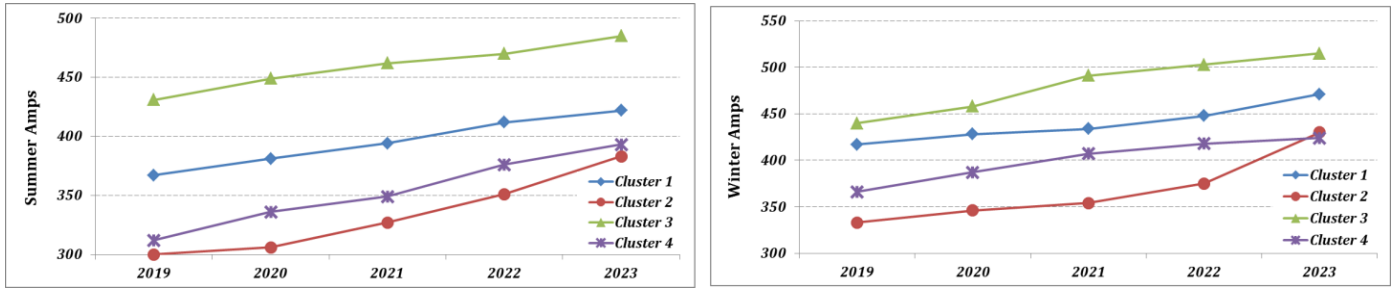

Figure 12. Average forecast results for the next 5 years by feeder clusters

Table 11: Application Resutls For Next 5 Years

| Feeder Cluster ID | Summer Average Growth Rate (%) | Winter Average Growth Rate (%) |
|---|---|---|
| 1 | 3.56 | 3.10 |
| 2 | 6.33 | 6.70 |
| 3 | 3.00 | 4.03 |
| 4 | 4.95 | 3.76 |

As Table 11 indicates, the growth of Cluster 2 (commercial heavy) is more rapid than the other 3 clusters due to more expected commercial load developments in the area in the forecasting years. For example, a major commercial development expected to energize in 2023 winter will increase the cluster loading significantly. The gathered customer net load changes that are going to occur at different time points are incorporated by the proposed model for different feeders and affect their loading growths over the forecasting years. Also, it is observed that the winter loadings are generally higher than the summer loadings across all clusters. This is because the tested planning area in Canada is a typical winter peaking system that has more electricity consumed in winter months for heating purpose and the same load can consume electricity differently between summer and winter seasons.

## 9. Conclusions and Discussions

This paper thoroughly discussed a novel method that can effectively forecast peak load of distribution feeders in an area over multiple years in the future. Compared to the commonly used top-down and bottom-up LTLF methods, as a hybrid forecasting method, the proposed method can seamlessly integrate top-down area features and bottom-up feeder features to improve forecasting accuracy; also, compared to the existing hybrid forecasting approach for area feeder LTLF, the proposed method:

- uses sequential learning with three different sequential configurations to convert a multi-year LTLF problem to a multi-timestep sequence prediction problem.

- uses a novel configuration selection mechanism that can learn, register and apply the best sequential configuration for each feeder in an area.
- uses unsupervised learning techniques to automatically group feeders into different groups and establish a set of sequence learning models for each group of feeders.

The proposed method was tested on a large urban distribution system in Canada. The results suggest:

- Learning and applying best-performing sequential configurations for different feeders accordingly provides better forecasting performance than applying single sequential configuration across all feeders in a planning area;
- Using clustering to help establish different learning models for different feeder groups in a planning area can improve forecasting performance;
- The proposed method successfully combines the top-down and bottom-up features in multiple historical years for multi-year ahead LTLF. It proves the feasibility of converting a multi-year LTLF problem to a multi-timestep sequence prediction problem.
- The proposed method outperforms various conventional methods and the existing hybrid forecasting method for the discussed problem.

In the future, when available, the proposed method should be tested on other area feeder datasets. The research can also be expanded to other types of component-level energy forecasting problems such as metering data forecast in a community. From the methodology improvement perspective, rather than using PCA for feature engineering, other non-linear dimensionality reduction methods such as T-SNE, auto-encoding as well as different feature selection methods can be considered and compared on the performance in the future.